# ESIA: An Efficient and Stable Identity Authentication for Internet of Vehicles

Haoxiang Luo, Jin Zhang, Xinling Li, Zonghang Li, Hongfang Yu,
Gang Sun, *Senior Member, IEEE*, Dusit Niyato, *Fellow, IEEE*

*Abstract*—Decentralized, tamper-proof blockchain is regarded as a solution to a challenging authentication issue in the Internet of Vehicles (IoVs). However, the consensus time and communication overhead of blockchain increase significantly as the number of vehicles connected to the blockchain. To address this issue, vehicular fog computing has been introduced to improve efficiency. However, existing studies ignore several key factors such as the number of vehicles in the fog computing system, which can impact the consensus communication overhead. Meanwhile, there is no comprehensive study on the stability of vehicular fog composition. The vehicle movement will lead to dynamic changes in fog. If the composition of vehicular fog is unstable, the blockchain formed by this fog computing system will be unstable, which can affect the consensus efficiency. With the above considerations, we propose an efficient and stable identity authentication (*ESIA*) empowered by hierarchical blockchain and fog computing. By grouping vehicles efficiently, *ESIA* has low communication complexity and achieves high stability. Moreover, to enhance the consensus security of the hierarchical blockchain, the consensus process is from the bottom layer to the up layer (bottom-up), which we call *B2UHChain*. Through theoretical analysis and simulation verification, our scheme achieves the design goals of high efficiency and stability while significantly improving the IoV scalability to the power of 1.5 (^1.5) under similar security to a single-layer blockchain. In addition, *ESIA* has less communication and computation overhead, lower latency, and higher throughput than other baseline authentication schemes.

## I. INTRODUCTION

In recent years, the rapid advancement of wireless communication technology has led to the emergence of the Internet of Vehicles (IoV) as a prominent research topic in the domains of communication and transportation. IoV is a vast interactive network that encompasses location, speed, and route information [1]. The transmission of data from sensors and other signal sources occurs through high bandwidth, low latency, and highly reliable links, thereby facilitating the realization of autonomous driving. The IoV communication technology enables vehicles to exchange information on road traffic and vehicle conditions. Upon receiving such information, other vehicles can adjust their driving routes to avoid potential traffic accidents and congestion, thereby enhancing vehicle traffic efficiency and safety. In [2], the author proposed an intersection-based distributed routing (IDR) strategy, which facilitates the dynamic formation of vehicle aggregation by the vehicles waiting at the intersection. These vehicles can actively establish multi-hop links with adjacent intersections, thereby effectively reducing the routing workload.

The proliferation of vehicles in the IoV has led to an exponential increase in the volume of information exchanged, exacerbating the communication and network security challenges inherent in the system. To establish a secure environment for IoV, identity authentication serves as the primary line of defense against malicious actors seeking to infiltrate the network. Given the criticality of ensuring the security and reliability of vehicle information collection, there is an urgent need for IoV security certification. Consequently, scholars have devoted considerable attention to this issue, seeking to enhance the security and reliability of IoV through various measures, including vehicle authentication [3]-[4]. This involves verifying the legitimacy of vehicles prior to their admission into IoV [5-6]. Recent advances in authentication methods have included the issuance of anonymous certificates by regional servers, which enable independent authentication of vehicle identity [7] and certificateless authentication methods [8-9]. However, these approaches often rely on centralized network topologies, which are vulnerable to traditional cryptography threats, such as single-point failure [10].

*A. Research Motivation*

Compared to traditional cryptographic algorithms, blockchain has demonstrated superior security performance

This research was supported by the Natural Science Foundation of Sichuan Province (2022NSFSC0913). (Corresponding author: *Hongfang Yu, Gang Sun*).

Haoxiang Luo, Jin Zhang, Zonghang Li, and Gang Sun are with the Key Laboratory of Optical Fiber Sensing and Communications (Ministry of Education), University of Electronic Science and Technology of China, Chengdu 611731, China (e-mail: lhx991115@163.com; zhangjin@uestc.edu.cn, lizhuestc@gmail.com; gangsun@uestc.edu.cn).

Xinling Li is with the Glasgow College, University of Electronic Science and Technology of China, Chengdu 611731, China, and also with the James Watt School of Engineering, University of Glasgow, Glasgow G12 8QQ, United Kingdom (e-mail: 2510870L@student.gla.ac.uk).

Hongfang Yu is with the Key Laboratory of Optical Fiber Sensing and Communications (Ministry of Education), University of Electronic Science and Technology of China, Chengdu 611731, China, and also with the Peng Cheng Laboratory, Shenzhen 518066, China (e-mail: yuhf@uestc.edu.cn).

Dusit Niyato is with the School of Computer Science and Engineering, Nanyang Technological University, Singapore 639798 (e-mail: dniyato@ntu.edu.sg).

and resolved many existing challenges [11-15]. As a result, blockchain has been identified as a promising solution to address the security challenges in IoV [16]. Consensus algorithms play a crucial role in ensuring network consistency in blockchain technology [17], with Practical Byzantine Fault Tolerance (PBFT) being one of the most widely used algorithms [18]. However, the communication complexity of PBFT increases polynomially with the number of nodes, which significantly affects blockchain scalability. This effect also exists in other consensus algorithms, such as Proof of Knowledge (PoK) [19]. Therefore, in the complex IoV with a large number of connected vehicles, it is necessary to improve consensus communication complexity to enhance the scalability and efficiency of the blockchain system.

Vehicular fog computing has emerged as a potential solution to address the above challenges associated with the large and uncertain scale IoV, as well as the potential latency caused by vehicle identity authentication [20-21]. This approach involves collecting, processing, and storing real-time traffic data by vehicles to improve communication efficiency [22]. In previous studies, vehicles have been divided into different fogs[1], and authentication has been carried out in each fog, resulting in a significant reduction in communication costs [21].

Currently, there are efforts underway to combine blockchain and fog computing to improve the efficiency and security of identity authentication [10], [23]. These efforts typically utilize vehicular fog computing to enhance authentication efficiency, and blockchain technology to ensure the security of each fog for recording information. However, there are two important issues that have not received sufficient attention in these studies. **1) There is a lack of literature that addresses the grouping scheme of vehicular fog from the perspective of minimizing communication complexity.** If each fog carries out blockchain consensus separately and parallelly, the consensus communication overhead will be closely related to the number of vehicles, making it essential to analyze the number of vehicles in each fog for efficiency purposes. **2) There are few studies on the stability of vehicle-fog composition in blockchain systems.** In our paper, we define stability as the composition of the fog, where the entry, exit, and renewal of vehicles can impact the fog stability (more details in Section VI-C). Each fog will reach a consensus in its own blockchain, and any changes to the vehicles in the fog can significantly impact the connection and communication of the blockchain, thereby affecting the efficiency and performance of the consensus. Therefore, the stability of the vehicle-fog composition will be closely related to various performances of the blockchain system, which warrants further investigation.

### B. Our Contributions

Capitalizing on the advantages of blockchain and fog computing, we propose a novel identity authentication scheme

---

[1] The "fog" and "fog computing system" are used interchangeably in our paper

---

for IoV called ESIA. This scheme involves the formation of a large number of fogs based on the driving and position of vehicles. Each fog selects a fog head vehicle (FV), while the other vehicles are referred to as ordinary vehicles (OVs). Each fog forms a blockchain to ensure the security of vehicle identity authentication, while all FVs form another separate blockchain to share vehicle information from each fog. This hierarchical structure forms a two-tier blockchain within the entire IoV, consisting of the ordinary vehicle layer (OVL) and fog head vehicle layer (FVL). This approach can enhance the scalability of IoV by limiting the consensus communication complexity within each fog, and can improve consensus efficiency by parallelizing the consensus of each fog. Our primary contributions are summarized as follows:

- Our proposed approach involves a hierarchical blockchain with low communication complexity and high stability. To ensure the security of the consensus process in this structure, the consensus process is bottom-up, leading to the structure being referred to as **B2UHChain**.
- Based on this hierarchical blockchain, we have designed an authentication scheme called **ESIA**, which involves four steps: initialization, registration, authentication, and logout. Compared to other authentication schemes, ESIA has lower communication and computation overhead, smaller latency, and higher throughput.
- To improve the consensus efficiency of IoV, we have designed a vehicular fog grouping scheme with low communication complexity, which quantifies the scalability of the hierarchical blockchain in IoV.
- To guarantee the stability of vehicular fog composition, we also offer the selection method of FV, and the composition method of each fog. To our best known, this is the first work exploring the stability of vehicle composition in the hierarchical blockchain.

### C. Structure of the Paper

The remaining contents of this paper are arranged as follows. Section II reviews the related work. Section III is the system model of our B2UHChain. Section IV describes the method of vehicle grouping, including fog grouping, fog head selection, fog composition, and dynamic change of grouping. The identity authentication process of ESIA is in Section V. Section VI is the security evaluation, including consensus security and ESIA security. In Section VII, the performance evaluation is presented in detail, including consensus communication complexity, IoV scalability, fog stability, storage overhead, communication and computation overhead, latency and throughput. Finally, Section VIII summarizes this paper.

## II. RELATED WORK

### A. Blockchain-enabled Identity Authentication in IoV

In the existing literature, researchers have explored various approaches to designing identity authentication schemes for IoV using blockchain technology. For instance, Jyoti [24] conducted a review of blockchain security solutions for IoV, emphasizing the crucial role of blockchain technology in identity scenarios. In [25], a certificateless authentication protocol was developed based on blockchain mechanisms and cryptography algorithms to achieve the security characteristics of open wireless communication in IoV. The protocol offers resistance against existential unforgeability in the face of chosen-message attacks. Similarly, in [26], the authors proposed an identity authentication scheme for IoV that leverages multiple semi-trusted institutions and the blockchain to create a distributed environment. The scheme employs certificateless signature and key distribution to achieve secure authentication with low computation and storage costs. In [27], the authors selected a leader vehicle from a group of vehicles to form multiple signatures for messages that are safely received by their member vehicles. These messages, along with their multiple signatures, are forwarded to RSUs near the leader vehicle, and then to a blockchain system maintained by P2P cloud servers.

In addition to cryptography-based methods that combine blockchain to design identity authentication for IoV, some scholars have also focused on improving consensus efficiency. For example, Xu *et al.* [16] proposed a secure and efficient distributed consensus algorithm called SG-PBFT for IoV identity authentication, which can enhance consensus efficiency and effectively prevent single-node attacks. Mershad *et al.* [28] developed a framework based on a private blockchain to secure vehicle-generated transactions. They designed a consensus mechanism called Proof of Accumulated (PoAT) and selected RSUs to act as blockchain miners.

For ESIA, it not only combines advanced cryptographic algorithms such as elliptic curve cryptography (ECC) and SHA-256, but also leverages the structure of the blockchain to minimize the consensus communication complexity and improve consensus efficiency.

### B. Fog-based Identity Authentication in IoV

The blockchain technology discussed above has proven to be an effective solution to the security problem in IoV identity authentication, while fog computing can address the communication efficiency issues in large-scale vehicular networks [29]. For instance, Song *et al.* [21] designed an IoV-oriented identity authentication scheme based on fog computing and ECC, which is resistant to various attacks. In [30], the authors proposed a charging identity authentication scheme based on fog computing to address the identity authentication problem in the vehicle-to-grid network. In this scheme, when electric vehicles need to charge, the relevant fog server must authenticate their identity and store the relevant electricity information. In [31], the authors developed a new certificateless signature scheme that employs a fog computing framework combined with online/offline encryption to reduce time consumption and improve authentication efficiency.

Furthermore, some researchers have explored the application of the combination of fog computing and blockchain in identity authentication, as demonstrated in [10], [23], and [32]. Although these studies have introduced fog computing to address communication latency issues, we have not come across any literature that focuses on minimizing communication overhead through fog grouping, such as determining the optimal number and size of fogs.

### C. Hierarchical Blockchain Systems

With the introduction of fog computing, OVL and FVL are formed, resulting in a hierarchical blockchain structure, and blockchain consensus takes place in these two layers. Therefore, it is crucial to investigate the existing layered blockchain methods.

Recent studies have proposed similar hierarchical blockchain systems. For instance, in [19], the authors proposed the PoK consensus with federated learning by dividing IoV into two layers to manage vehicles and enhance vehicle scalability. In [33], the authors allocated nodes participating in consensus to specific layers and propose an improved consensus algorithm. In [34], RSUs are taken as layered objects and combined with blockchain to achieve the rapid revocation of certificates in IoV. Luo *et al.* [35] proposed two hierarchical consensus algorithms, PRAFT and RPBFT, that improve blockchain scalability. Cui *et al.* [36] divided nodes into three layers based on their functions and propose an identity authentication scheme.

Through the literature review, we find that few studies have focused on the node allocation strategy in hierarchical blockchain, specifically, which nodes should be in which layer. The dynamic node allocation strategy for blockchain is essential in the vehicular mobility environment because we need to determine which vehicles should be in a particular fog. In this regard, Cheng *et al.* [37] grouped vehicles based on their attributes, such as direction and location. In [38], the authors designed a diffused PBFT (DPBFT) consensus for IoV identity authentication to reach consensus in the network space, which can reduce consensus overhead. However, these methods do not analyze the number of vehicles in each fog, which is crucial for consensus efficiency. Furthermore, they do not consider the stability of vehicular fog composition. If the constituent vehicles of fog are unstable, the blockchain of OVL and FVL will also be unstable, which can affect vehicle connection and consensus efficiency.

### III. SYSTEM MODEL OF B2UHCHAIN

Blockchain can enhance security in the authentication process of IoV. However, the traditional blockchain structure may not be suitable when the number of vehicles is large. To address this issue, B2UHChain divides vehicles into the OVL and the FVL. Then, the consensus process is carried out separately and parallelly to reduce communication complexity and increase the IoV scalability. FVs are also included in the consensus process in the OVL. Once the FVL has reached consistency, it sends the authentication information to the RSUs layer, which does not involve the blockchain. The B2UHChain structure is depicted in Fig. 1, and the network topology of information transmission is presented in Fig. 2.

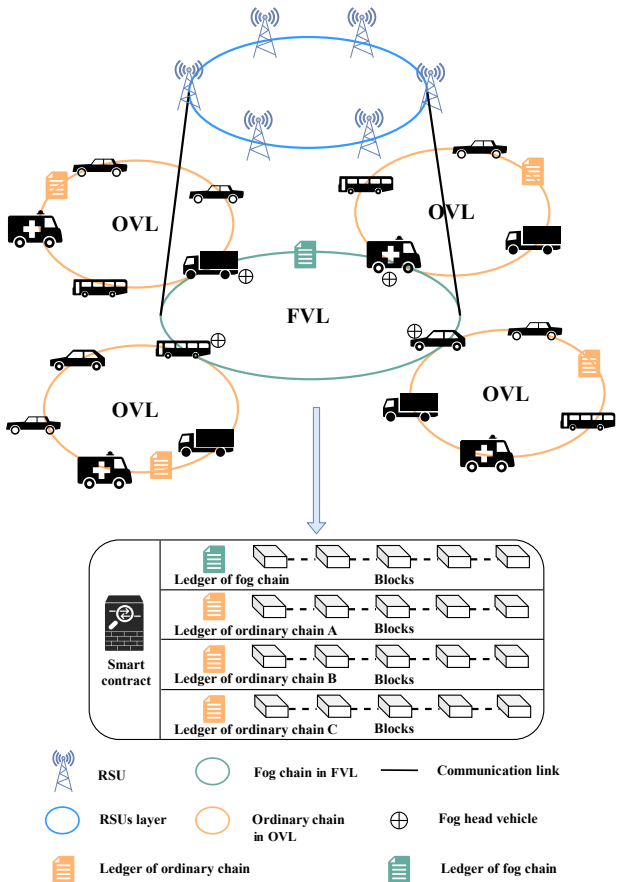

Fig. 1. B2UHChain structure.

### A. Ordinary Vehicle Layer

The OVL is composed of several distinct patches of fog, with each fog containing a different set of OVs and a selected FV forming a separate ordinary chain. The blockchain consensus process for identity authentication is carried out simultaneously in each fog on the OVL. It is essential to note that OVs are only permitted to communicate within their respective fog. Therefore, the consensus result of each fog is stored in its corresponding blockchain ledger, and each fog must maintain its own ledger. If an OV wishes to communicate with a vehicle in another fog, the FVs in these two fogs must carry out identity authentication and forward the OV's information that needs to communicate. Moreover, OVs must be registered and identified in the smart contract.

### B. Fog Head Vehicle Layer

The FVL comprises FVs selected from each fog, which form a fog chain. After the consensus process of the OVL, the FV has the consensus result and authentication information from the fog, and then it carries out the consensus of identity authentication with other FVs in the FVL. Each FV can communicate with OVs in the same fog, other FVs, and RSUs. It can also receive and forward various data from OVs to RSUs, and can access vehicle state data in RSUs. Similarly, the consensus result of the fog chain is stored in its corresponding ledger, and the fog chain also maintains it. Further, FVs must be registered and identified in the smart contract.

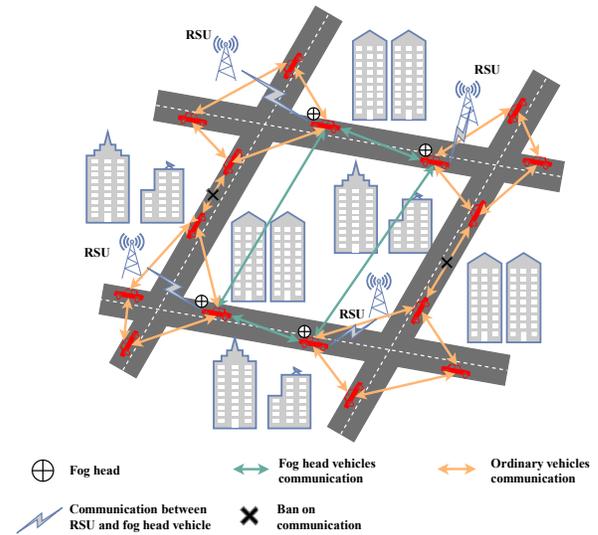

Fig. 2. Network topology of information transmission in B2UHChain.

### C. RSU Layer

The RSU layer comprises roadside units that manage vehicles in the IoV, receive data from FVs, and process, store, and analyze this data [39]. Since RSUs are typically run by government departments or state-owned enterprises in most countries, such as China, they can be assumed to be a trusted authority (*TA*) [40]. RSUs act as vehicle managers and are responsible for generating system secret keys, public keys, and control parameters, and preloading them to the corresponding vehicles. This means that vehicles must be initialized by the RSUs before joining the network. Additionally, RSUs can track vehicles based on their anonymous identity and determine FVs based on vehicle information [40].

This paper focuses primarily on the consensus efficiency and security of blockchain-based identity authentication between vehicles, and does not consider the communication security between RSUs and vehicles or between different RSUs, because this part has been already studied extensively [3, 38], and it is beyond the scope of our paper. We assume that these communications are secure and do not form a blockchain like OVL and FVL do.

## IV. VEHICLES GROUPING FOR B2UHCHAIN

This section provides a detailed discussion of the number of fogs, the number of vehicles in each fog, and the vehicle identity, specifically whether it is an OV or an FV.

### A. Number of Vehicles

This part delves into the optimal number of fogs and vehicles in each fog. Due to the high requirements for transaction processing efficiency in the IoV, it is necessary to reduce the communication complexity of blockchain when reaching consensus. The communication complexity of the PBFT consensus algorithm often depends on the number of vehicles in the chain. Therefore, we propose a layered blockchain structure based on different vehicle attributes and functions to limit the communication of vehicles within each fog, thus reducing the communication complexity needed to reach a consensus and improving the IoV scalability.

Unlike the multi-layer scheme in [33] and PRAFT in [35], which are top-down schemes with a low consensus success rate, as demonstrated in Section VI-A, our B2UHChain is a bottom-up process to ensure the consensus security. That is, the consensus of the OVL is carried out first, followed by the consensus of the FVL. Combined with the characteristics of the number of nodes in PBFT, we set the number of vehicles in each fog to not be less than four [18, 35].

We set the number of all vehicles in the IoV as $Z$, and then the communication complexity of using the PBFT consensus in the traditional single-layer blockchain structure is the number of nodes squared $O(Z^2)$ [18]. First, we analyze the situation of the same number of vehicles in each fog after dividing the vehicles into two layers. If there are $x$ FVs in the FVL, then there are $x$ fogs in the OVL, and we have $y$ OVs in each fog. Therefore, $x$, $y$, and $Z$ satisfy the following relation

$$Z = x + xy, \quad (1)$$

$$x = \frac{Z}{1+y}. \quad (2)$$

The communication complexity $C$ of the B2UHChain reaching consensus is (3), and the definition of communication complexity is same of [33] and [35], which is the communication overhead of the whole network.

$$C = x^2 + xy^2. \quad (3)$$

By combining (2) and (3), we can get

$$C = \left(\frac{Z}{1+y}\right)^2 + \frac{y^2 Z}{1+y}. \quad (4)$$

The first and second derivatives of $C$ can be obtained by taking $C$ as a function of $y$

$$\frac{dC}{dy} = \frac{-2Z^2}{(1+y)^3} + \frac{Zy(y+2)}{(1+y)^2}, \quad (5)$$

$$\frac{d^2C}{dy^2} = \frac{6Z^2}{(1+y)^6} + \frac{2Z}{(1+y)^3}. \quad (6)$$

As $x$, $y$, and $Z$ are constrained to positive integers, it can be inferred that $d^2C/dy^2$ is necessarily greater than zero. This implies that $dC/dy$, being a monotonically increasing function, also exists a zero value. As a result, there is a minimum for $C(y)$. Let the value of (5) is equal to 0, and we can get

$$Z = \frac{y^3 + 3y^2 + 2y}{2}, \left(4 \leq y \leq \frac{Z-4}{4}\right). \quad (7)$$

Then, $C(y)$ is

$$C(y) = \frac{3y^4 + 8y^3 + 4y^2}{4}. \quad (8)$$

We divide (8) by (7)

$$\frac{C(y)}{Z} = \frac{3y^3 + 8y^2 + 4y}{2y^2 + 6y + 8} \approx \frac{3y-1}{2}. \quad (9)$$

Combining (1), (7), and (8), the minimum value of communication complexity to reach PBFT consensus in B2UHChain is

$$C = 1.89 Z^{\frac{4}{3}}. \quad (10)$$

Thus, by layering IoV, we reduce the communication complexity of the blockchain consensus from the square to the power of 4/3. According to (2) and (7), when the communication complexity is the minimum, $x$ and $y$ meet the requirement

$$x = \frac{y(y+2)}{2}. \quad (11)$$

When the calculated result of (11) is a decimal, similar to [35], we round it down to an integer. Additionally, in the event of an uneven distribution of vehicles across fog, the resulting communication complexity is higher than that of evenly distributed, as predicted by the mean value inequality.

### B. Fog Head Selection

This part pertains to the determination of which vehicles are classified as FVs. In the B2UHChain, FVs are responsible for collecting information regarding OVs within the same fog, as well as engaging in consensus with other FVs. It is worth noting that the effectiveness of Vehicle-to-Vehicle (V2V) communication in the IoV is often influenced by factors such as communication protocols, the distance between vehicles, and their relative speed. As such, the appropriate selection of FVs plays a pivotal role to the efficiency and latency of blockchain consensus within the fog.

The position, direction, and speed of each vehicle are critical parameters in V2V communication, which can be accurately determined in real-time through GPS systems while the vehicle is in motion. Additionally, the vehicle is equipped with knowledge of its own route and distance. By exchanging this information with nearby vehicles, the vehicle can calculate the number of adjacent vehicles and their respective distances. Based on the above assumptions, for the $i$th vehicle, we define the fog head parameter $\alpha_i$ as (12), where $n_i$ denotes the number of vehicles around the $i$th vehicle, and $s_i$ is the distance it wants to travel on this road.

$$\alpha_i = l \cdot \frac{1}{s_i} + m \cdot \frac{1}{n_i}. \quad (12)$$

where, $l$ and $m$ are two parameters that satisfy $l+m=1$. The physical meaning of $\alpha_i$ is the closeness of the $i$th vehicle to its neighbors on this road. The smaller the fog head factor $\alpha_i$ is, the more vehicles can be covered by the blockchain formed within the communication range of the vehicle, and the more stable the vehicle's running state is. As a result, the vehicle is more likely to become the FV. Once the grouping of vehicles is determined, the vehicle with the smallest fog head factors in each fog can be selected as the FV.

### C. Fog Composition

This part pertains to the selection of vehicles to be included in each fog. The efficiency of fog consensus is influenced by several factors such as the distance between vehicles, their relative speed, and communication range. Therefore, it is essential to devise a suitable scheme for vehicle composition within the fog to ensure optimal efficiency of the consensus process. We define the correlation degree between the $i$th vehicle and other vehicles in the fog as the fog factor $\beta_i$,

| | |
|---|---|
| **Algorithm 1: Method of Vehicle Grouping** | |
| 1 | **Initialization:** Count the total number of vehicles $Z$ |
| 2 | **Fog Number executes:** |
| 3 | Compute $x$ and $y$ as equations (1) and (11) |
| 4 | **FV Selection executes:** |
| 5 | **Input:** $s_i$, $n_i$, $l$ and $m$ |
| 6 | **for** $1 \leq i \leq Z$, **do** |
| 7 | Compute $\alpha_i$ as equation (12) |
| 8 | Choose $\alpha_{\min} = \min(\alpha_i)$ |
| 9 | **end for** |
| 10 | **Output:** $x$ vehicles with the smallest $\alpha_i$ value are selected as FV |
| 11 | **Fog Composition executes:** |
| 12 | **Input:** $v_i$, $\bar{v}$, $d_i$, $s_i$, $S$, $a$, $b$ and $c$ |
| 13 | Compute $\beta_i$ as equation (13) for each OV |
| 14 | **for** $\beta_{th}$ =0:0.1:1, **then** |
| 15 | **if** $\beta_i < \beta_{th}$, **then** |
| 16 | The $i$th OV can join this fog |
| 17 | **else** the $i$th OV cannot join this fog, and selects another fog to re-evaluate the $\beta_i$ and $\beta_{th}$ values |
| 18 | **end if** |
| 19 | **if** this $\beta_i$ value is the lowest value for the $i$th OV, **then** |
| 20 | **if** the OV number $< y$ in this fog, **then** |
| 21 | The $i$th OV joins this fog |
| 22 | **else if** there are OVs whose $\beta_i$ values are greater than that of the $i$th OV, **then** |
| 23 | The fog removes the OV with the maximum $\beta_i$ value and adds this OV |
| 24 | **else** reject the $i$th OV to join, and this OV chooses another fog that satisfies $\beta_i < \beta_{th}$ |
| 25 | **end if** |
| 26 | **else** the $i$th OV chooses another fog when it has the lowest $\beta_i$ value to repeat the lines 20-25 |
| 27 | **end if** |
| 28 | **if** $x$ and $y$ satisfy equation (11), **do** |
| 29 | Output this $\beta_{th}$ with its fog composition |
| 30 | **end if** |
| 31 | **end for** |
| 32 | **Output:** the composition of each fog |

which is related to the speed of vehicle $v_i$, the average speed $\bar{v}$ of vehicles in this fog, the distance between the vehicle and FV $d_i$, and the vehicle communication distance $S$:

$$\beta_i = a \cdot \left| \frac{v_i - \bar{v}}{\bar{v}} \right| + b \cdot \frac{1}{s_i} + c \cdot \frac{d_i}{S} \quad (13)$$

where $a$, $b$, and $c$ are all parameters, satisfying $a+b+c=1$. Similar to the $\alpha_i$, the smaller the $\beta_i$ value, the higher the correlation degree between vehicles, and vehicles are more likely to join this fog naturally. Each vehicle has a different $\beta_i$ value for different fogs.

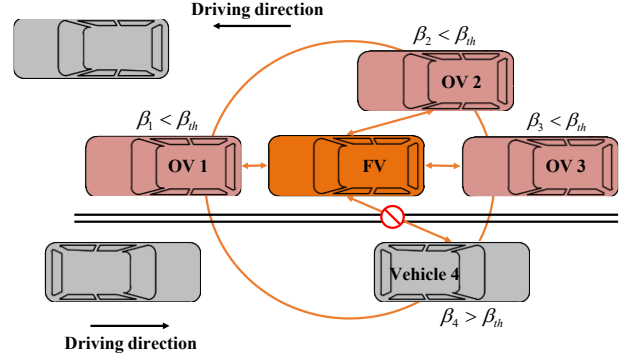

Fig. 3. Fog diagram.

Moreover, we have established a uniform threshold for the fog factor $\beta_{th}$ in each fog. Only vehicles whose fog factor is less than the set threshold can be included in the respective fog. In cases where a vehicle is eligible for inclusion in multiple fogs, it will select the fog when it has the lowest $\beta_i$ value. In addition, when the number of vehicles in the fog added by the OV has reached the upper limit, the FV will eliminate the vehicles with the largest $\beta_i$ value. If the vehicles are not evenly distributed within each fog, the excess vehicles will choose to join the fog when it has the lowest $\beta_i$ value.

According to the $\beta_{th}$ value set, a smaller $\beta_{th}$ would result in higher consistency among vehicles within the fog, leading to faster consensus processes. However, the number of eligible vehicles would decrease, resulting in an increase in the number of fogs. Through many preliminary simulations (same as Section VII-D), it is easier for us to obtain the situation satisfying (1) and (11) by iterating $\beta_{th}$ values from 0 to 1. Then, the communication complexity can be reduced to the minimum value (10).

Fig. 3 shows whether OV is added to the fog. To summarize, **Algorithm 1** is the whole fog grouping algorithm.

### D. Time complexity analysis

The IoV is a dynamic and time-varying environment given that vehicles are constantly in motion. To determine the applicability of Algorithm 1 for B2UHChain in IoV, it is necessary to analyze its time complexity. This can be accomplished by evaluating three parts of the algorithm.

First, considering the number of fogs and vehicles in each fog (Part *A* of this Section), which corresponds to lines 2-3 in Algorithm 1, the time complexity is constant and can be expressed as $O(1)$.

Second, for FV selections (Part *B* of this Section), which involves the calculation of the fog head parameter $\alpha_i$ for each vehicle (lines 4-10), the time complexity is linear and can be expressed as $O(Z)$, where $Z$ denotes the number of vehicles in the IoV.

Finally, when analyzing the fog composition (Part *C* of this Section), which involves lines 11-32 in the algorithm. This portion contains a "for" loop from 0 to 0.1 and five "if" statements within the loop, four of which pertain to individual vehicles while the last one concerns the B2UHChain system. Thus, given $Z$ vehicles, the time complexity for this section is $40Z + 10$ and can be expressed as $O(Z)$.

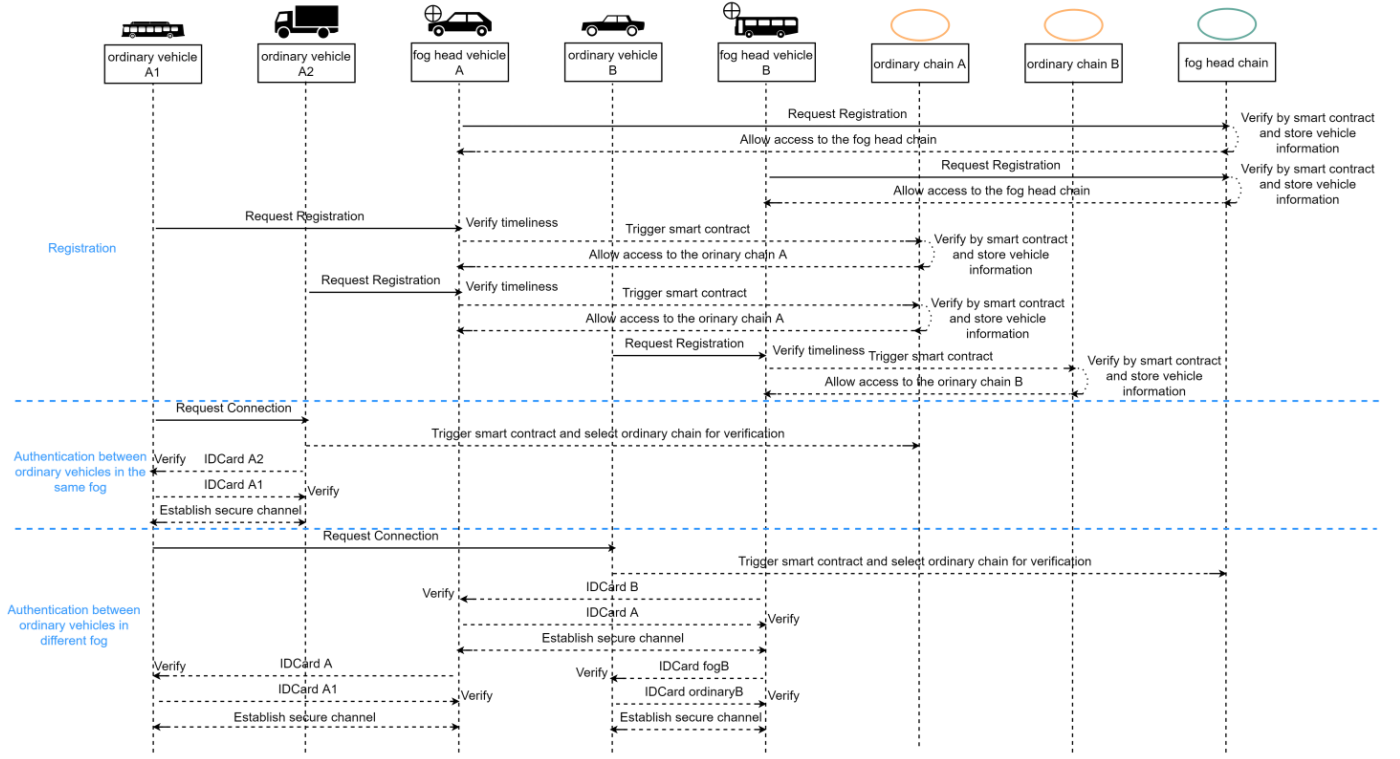

Fig. 4. Overall flow of the ESIA scheme.

Overall, Algorithm 1 displays low time complexity at the linear level of $O(Z)$ and is therefore appropriate for application in the time-varying IoV environment. In the following part, we will discuss the dynamic changes that trigger the restart of Algorithm 1.

*E. Dynamic Changes of Grouping*

The dynamic nature of vehicle states necessitates a discussion of the grouping method following changes in a vehicle's state. This part considers several cases of dynamic changes in the grouping.

*1) The total number of vehicles Z has changed.* The total number of vehicles $Z$ will change when vehicles leave or some vehicles join the group. In such cases, B2UHChain must re-evaluate the fog grouping method based on the updated $Z$ value, as defined by equations (1) and (11).

*2) The FV has changed.* The identity of an FV may change when there are significant changes in its speed and position. In such cases, the FV may become an OV, and a new FV must be identified in accordance with equation (12). Conversely, an OV can also replace the old FV and become a new fog head.

*3) The OV has changed.* There are three variations of the OV identity. The first one is that it becomes an FV; Second, the FV may change, and its driving characteristics, as defined by equation (13), no longer meet the requirements of the new fog, necessitating its entry into another fog; Third, the fog in which the OV is located has not changed, but its driving characteristics have changed significantly according to (13). In this case, its $\beta_i$ value is bigger than the $\beta_{th}$ value of this fog, thus, the OV no longer meets the requirements of this fog, and will be removed by the FV. Then, it is required to enter a new fog.

## V. ESIA PROCESS

Drawing from the B2UHChain framework described earlier, this section designs an identity authentication process. The authentication process is divided into four distinct steps, with the primary processes of registration and authentication depicted in Fig. 4.

*1) Initialization:* Initialize the information and parameters of all vehicles in IoV through RSU;

*2) Registration:* The identity of each vehicle is registered and stored on the blockchain;

*3) Authentication:* Identity authentication requests for different types of vehicles are validated and authorized through a B2UHChain model of the IoV;

*4) Logout:* When the vehicle is out of the communication range of the FV, it needs to be logged out in the fog.

*A. Initialization*

First, we initialize all vehicles in IoVs through RSUs. The RSU generates the identity of each vehicle, whether it is an OV or an FV. Since each vehicle has a unique Ethernet address (EA) in the network [36], we can transform the Ethernet address using the hash function through RSU, getting the unique identity of the vehicle $ID_i=Hash(EA_i)$, and then send it to the vehicle for storage. Among them, the identity of OV is *OID*, the identity of FV is *FogID*, and the identity of RSU is *RSUID*.

Second, we set $u \in Z_q$ and $v \in Z_q$ be two constants, where $Z_q=\{0, 1, …, q-1\}$, ($q>3$) should be prime. Then, the *TA* chooses "a non-singular elliptic curve $E_q(o, p)$: $u^2 =v^3+ov+p$ over the finite field $GF(q)$, such that $4v^3+27u^2 \neq 0 \pmod q$ with a point at infinity or zero point $O$". Suppose $G$ is taken as the base point in $E_q(o, p)$ and its order is same as $q$. Then $TA$ generates a private key $sk_i \in Z_q^*$ and its corresponding public key $pk_i \in sk_iG$ for each vehicle by ECC. The public and private keys generated here are primarily used to verify the integrity of messages sent during registration and authentication.

Finally, the $TA$ also generates an $IDCard$ for each vehicle to prove its unique identity. The $IDCard$ structure of the FV is $sig(Hash(RSUID\|FogID))$, namely the signature value of $RSUID$ and the vehicle's own identity $FogID$; The $IDCard$ structure of the OV is $sig(Hash(RSUID\|FogID\|OID))$, namely the signature value of $RSUID$, the according FV identity $FogID$, and the vehicle's own identity $OID$. The signature result contains two parts $R$ and $s$, where $R$ represents the result of a random number $r$ obtained by ECC, and $s$ represents a calculated intermediate value [11]. The specific calculation process is as follows

$$R = rG,$$
$$s = (h + sk_i R_x)/r, \quad (14)$$

where the $h$ is abbreviation of the hash value $Hash(\cdot)$, and $R_x$ denotes the value of $R$ on the $X$-axis.

### B. Registration

In the two-layer IoV forming, the identity information of the OV is bound to its corresponding FV, and the identity information of the FV is bound to the RSU, both of which are stored on the blockchain. We define $Time$ and $Time^*$ indicate the timestamp of the transaction created and the vehicle received the transaction, respectively. And the $\Delta TS$ represents the allowed time interval for transactions to be sent.

Furthermore, the registration step consists of two steps: registration of FVs and OVs. After selecting FVs through Section IV, FVs are first registered in the blockchain, and then OVs are registered in the fog.

*1) FV Registration*

First, FV submits the request message $ReqReg(FogID, RSUID, IDCard_{FogID}, Time)$. Then it would follow the steps below to start the smart contract on the blockchain and perform the registration verification process:
- Verify whether the request timed out by $Time^*-Time \leqslant \Delta TS$.
- Verify the validity of the RSU identity $RSUID$.
- Use the $FogID$ and $RSUID$ in the request message $ReqReg$ and the public key from $TA$ to verify the correctness of the FV identity card $IDCard_{FogID}$. The verification process is as follows

$$R' = (hG + R_x pk_i)/s. \quad (15)$$

If $R=R'$, the verification is successful.

In the event that the previous verification steps are not all completed successfully, the FV will not be registered, and a registration *error* message will be returned. Conversely, if all the verification processes are successfully completed, the blockchain will store the identity information of the FV in the designated format and publish the authenticated message. This indicates that the FV is authorized to access the IoV.

*2) OV Registration*

The registration and verification process for OV occurs on the blockchain when the fog is formed. Once the fog is selected according to the rules outlined in Section IV, the OV must broadcast its registration request message, $ReqReg(OID, FogID, RSUID, IDCard_{OID}, Time)$.

Upon receiving ReqReg, the FV will first verify the timeliness of the $Time$ parameter. If the timestamp is deemed valid, the transaction will be registered on the fog chain, and the smart contract responsible for the registration process of OV identity information will be initiated in the following sequence:
- Verify whether the request timed out by $Time^*-Time \leqslant \Delta TS$.
- Verify the validity of RSU identity $RSUID$, and the validity of the FV's identity $FogID$.
- Use the $OID$, $FogID$, and $RSUID$ in the request message $ReqReg$ and the public key from $TA$ to verify the correctness of the OV identity card $IDCard_{OID}$. The verification process is the same as the FV, that is, using the received $R$ and $s$, to calculate $R'$ and determine whether $R=R'$ is true.

Similar to FV registration, if the previous verification steps are not all completed successfully, the OV will not be registered, and a registration *error* message will be returned. Conversely, if all the verification processes are successfully completed, the blockchain will store the identity information of the OV in the designated format and publish the authenticated message. This indicates that the OV is authorized to access the ordinary chain.

### C. Authentication

It should be noted that the vehicle's identity authentication process is the same as the PBFT consensus process mentioned above. The authentication is performed in the OVL first, and then in the FVL. However, the authentication process of OV and FV is almost the same, so only the authentication process of OV is introduced here.

Bidirectional authentication between OV $A1$ and OV $A2$ is initiated when they need to establish a secure communication channel. OV $A1$ first sends the connection request message $ReqCon(A1_{OID}, A1_{FogID}, A1_{RSUID}, A2_{OID}, IDCard_{A1})$. The smart contract is triggered when the OV $A2$ receives $ReqCon$ and verifies the authentication request on the fog chain with the following steps:
- According to the vehicle information stored in the blockchain, check whether the $IDCard_{A1}$ format held by the vehicle is correct. If the validation is successful, proceed with the following steps, otherwise, return the *error*.
- Search the identity information of vehicle $A1$ and vehicle $A2$ according to the vehicle identity information stored in the fog chain. An *error* is returned if vehicle $A1$ or $A2$ does not exist.
- Verify the status of vehicles $A1_{OID}$ and $A2_{OID}$, and returns

an *error* if they are not valid.
- Fog chain searches the identity information of vehicles $A1$ and $A2$ according to the *IDCard* information of vehicles stored in the blockchain, because the *IDCard* can show not only its identity information, but also fog information in which the vehicle is located. If they belong to the same fog, they return *true* directly to the corresponding FV, and vehicles $A1$ and $A2$ establish secure communication channels.

When vehicles $A$ and $B$ belong to different fogs, they must first authenticate with their respective FVs in the fog. Subsequently, the two FVs will perform mutual authentication in the FVL. Finally, the FVs will inform the authentication results of the two OVs. It is worth noting that the authentication process between FV and OV within the same fog is consistent with that of OV in the same fog. A detailed illustration of the authentication process for vehicles in the same and different fog is presented in Fig.5.

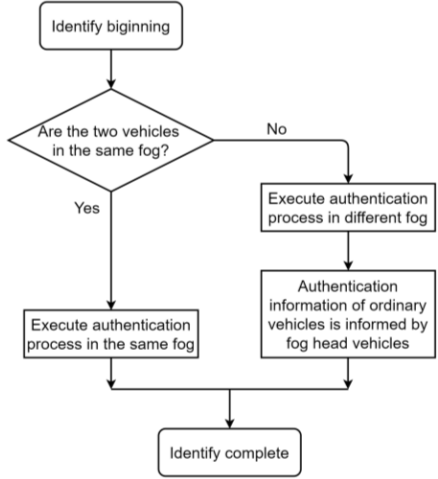

Fig. 5. Authentication process.

### D. Logout

FV submit information about their OV in fog that need to be logged out. And then, the FV sends a request message *ReqLogout*(*OID, FogID, RSUID*) to the fog chain. Furthermore, the smart contract executes the vehicle cancellation procedure to verify the existence of the RSU *RSUID*, the FV *FogID*, and the OV *OID* in turn. After the transaction verification is submitted, the *OID* of the OV will be revoked. When the FV changes, new fog needs to be formed, and a new round of vehicle authentication should be carried out from initialization.

## VI. SECURITY EVALUATION

This section analyzes the consensus security, and the ESIA security.

### A. Consensus Security
#### 1) Comparison with Single-layer PBFT

Same as [33] and [35], the consensus security is the consensus success rate. In this part, from the perspective of probability, we compare the security changes of B2UHChain and single-layer blockchain during PBFT. This consensus algorithm can accommodate the 1/3 faulty vehicles, that is, it needs 2/3 or more vehicles to reach consistency.

If the probability of failure of each vehicle is $P_f$, then the probability of reaching blockchain consensus under the single-layer blockchain is

$$Ps_1 = \sum_{i=0}^{\lfloor \frac{Z}{3} \rfloor} C_Z^i \left(1-P_f\right)^{(Z-i)} P_f^i. \tag{16}$$

In the B2UHChain consensus, the consensus of the OVL is first conducted, and the consensus success rate of each fog in this layer is

$$P_1 = \sum_{i=0}^{\lfloor \frac{y}{3} \rfloor} C_y^i \left(1-P_f\right)^{(y-i)} P_f^i. \tag{17}$$

The precondition for the FVL to reach consensus is that at least 2/3 ordinary fog in the OVL reach consensus, that is, less than 1/3 ordinary fog consensus fails. We set this event called Event $A$, then the probability of it occurring is

$$P(A) = \sum_{j=0}^{\lfloor \frac{x}{3} \rfloor} C_x^j \left(1-P_1\right)^j P_1^{(x-j)}. \tag{18}$$

Further, for FVL, it requires less than 1/3 faulty FVs to successfully reach consistency. Then, we call it Event $B$. If Event $A$ occurs, the probability of Event $B$ is

$$P(B \mid A) = \sum_{k=0}^{\lfloor \frac{x}{3} \rfloor - j} C_{x-j}^k \left(1-P_f\right)^{(x-k-j)} P_f^k. \tag{19}$$

Therefore, the consensus success rate of the B2UHChain is

$$\begin{aligned} Ps_2 &= P(AB) = P(A)P(B \mid A) \\ &= \sum_{j=0}^{\lfloor \frac{x}{3} \rfloor} C_x^j \left(1-P_1\right)^j P_1^{(x-j)} \sum_{k=0}^{\lfloor \frac{x}{3} \rfloor - j} C_{x-j}^k \left(1-P_f\right)^{(x-k-j)} P_f^k. \end{aligned} \tag{20}$$

Then, the consensus success rate change between B2UHChain and single-layer blockchains can be expressed by the following formula

$$I = \frac{Ps_2 - Ps_1}{Ps_1}. \tag{21}$$

The findings presented in Figs. 6 (a) and (b) are obtained through simulations utilizing varying values of the total number of vehicles, as dictated by the grouping rules (11). It is observed that when the failure rate of a vehicle exceeds 0.4, the B2UHChain exhibits higher consensus security, thereby rendering it more suitable for deployment in poor network conditions. Furthermore, as the value of $Z$ increases, the corresponding value of $I$ exhibits a more significant increase. Conversely, when the failure rate of a vehicle is relatively low (i.e., approximately 0-0.35), the security of the system is negatively impacted, and the value of $I$ decreases. A potential solution to this issue is presented in Section VII-C. Overall, the B2UHChain demonstrated a higher consensus success compared to the single-layer structure in most scenarios.

#### 2) Comparison with Other Hierarchical Blockchains

In this part, we compare the consensus security of

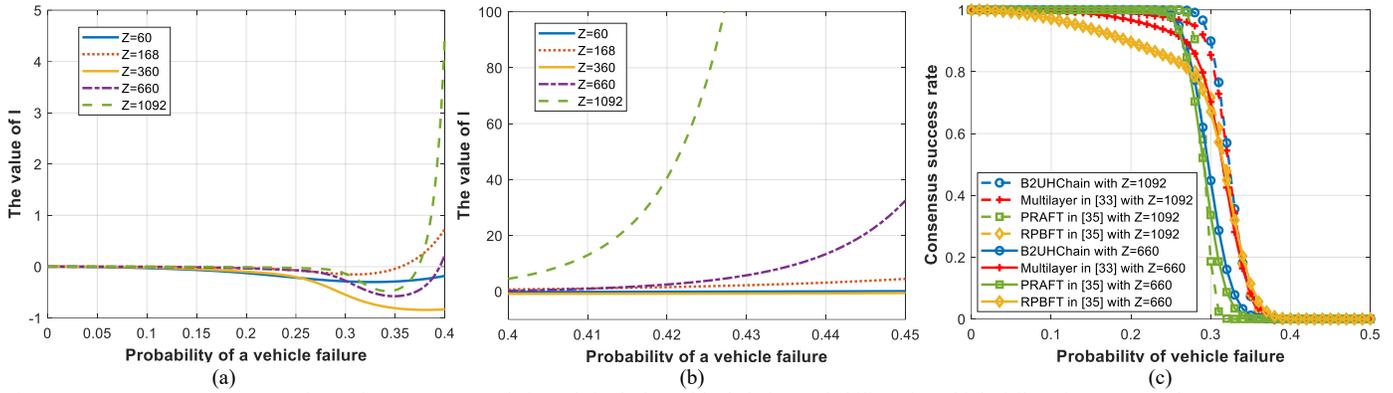

Fig. 6 (a) Consensus success rate change between B2UHChain and single-layer blockchain (Probability of a vehicle failure from 0-0.4). (b) Consensus success rate change between B2UHChain and single-layer blockchain (Probability of a vehicle failure from 0.4-0.45). (c) Consensus success rate of different hierarchical blockchains.

B2UHChain with that of other two-tier blockchains, such as the top-down scheme in [33], and PRAFT and RPBFT in [35]. The simulation result is shown in Fig. 6 (c). Our B2UHChain has higher consensus security than PRAFT and RPBFT in [35]. Meanwhile, in most cases, i.e., the probability of vehicle failure is small, B2UHChain is safer than the scheme in [33].

*B. ESIA Security*

This section pertains to the examination of the security and resistance of the ESIA against various attack methods. The aim of this analysis is to assess the scheme's ability to evaluate its overall robustness in the face of adversarial threats.

**Denial of Service Attack**. Since the fog chain in ESIA is composed of FVs, and the identity of each FV is generated by the RSU, it does not allow access to any illegal vehicles, making it difficult for attackers to directly attack the fog chain by this attack. Additionally, the ordinary chain requires a certain number of resources for each transaction submission, which makes it challenging for attackers to overload the ordinary chain by sending a large number of authentication requests.

**Double Spending Attack**. This attack is common in blockchain systems. For example, in Bitcoin, a malicious node can launch a double spending attack if it has more than 50% computing power of the Bitcoin network. However, in ESIA, the large scale of vehicles makes it difficult for attackers to control thousands of vehicular fog computing power to launch a double spending attack.

**Forgery Attack**. Each OV in ESIA has a unique identity *OID* in the IoV, which is linked to its RSU and FV corresponding to the unique *RSUID* and *FogID*. This makes it impossible for attackers to fake legitimate vehicles in the network to communicate with other vehicles.

**Message Replay Attack**. ESIA attaches a timestamp *Time* to the request message during the registration phase, which helps to avoid message replay attacks by determining the timeliness of the message. In the authentication phase, mutual authentication is required between two vehicles, which means that the replay attack of one vehicle cannot be authenticated successfully.

**Man in the Middle Attack**. Assuming that the attacker intercepts the information transmitted in the authentication process and uses the illegal vehicle to carry out this attack. In the registration phase, if an attacker obtains the registration message, but the illegal vehicle cannot access the IoV, the subsequent process of identity authentication cannot be completed. In the authentication phase, even if the attacker obtains the authentication message, the two vehicles need to recognize each other, so the illegal vehicle cannot complete the authentication through the obtained authentication message. Therefore, ESIA is resistant to this attack.

TABLE I
PARAMETERS OF SIMULATION

| Parameters | Values |
|---|---|
| Covered area | 2000*1600 $m^2$ |
| Simulation duration | 10 s |
| Number of traffic lanes | 2 |
| Number of traffic roads | 5*5 |
| MAC layer protocol | IEEE 802.11p |
| Communication distance of a vehicle | 150 m |

VII. PERFORMANCE EVALUATION

This section presents a comprehensive performance evaluation of the B2UHChain, as well as the different performances of ESIA based on this two-tier structure.

*A. Simulation Setup*

We use SUMO, an open-source traffic simulation software developed in C++, to simulate the vehicle flow of B2UHChain. It allows the simulation of a traffic environment consisting of multiple vehicles moving on a given traffic path. Table I shows the simulation parameters and simulation scenarios of this model.

*B. Consensus Communication Complexity*

Due to the rapid movement of vehicles, IoV has high requirements for consensus efficiency. Thus, we limit the communication overhead of PBFT to each fog in the OVL.

First, to evaluate the effectiveness of our proposed approach, we compare the consensus complexity of our B2UHChain with the single-layer PBFT, SG-PBFT [16], and the hierarchical blockchain Pyramid [41], as depicted in Fig. 7 (a). The results showed that the communication complexity of B2UHChain is significantly lower than that of single-layer PBFT and SG-PBFT, and even lower than that of the hierarchical scheme Pyramid.

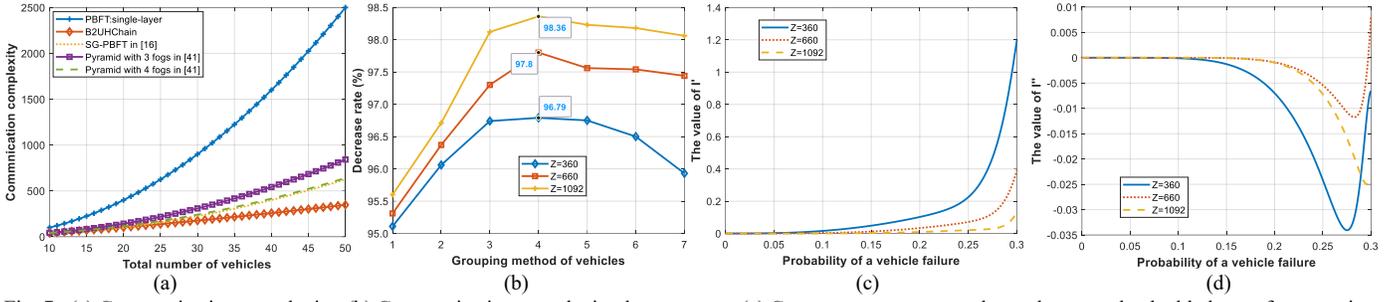

Fig. 7. (a) Communication complexity. (b) Communication complexity decrease rate. (c) Consensus success rate change between the double layer after capacity expansion and that before. (d) Consensus success rate change between the double layer after capacity expansion and the single layer.

Second, we utilize equations (1) and (11) to set the values of $y=8$, $x=40$, and $Z=360$, while keeping the total number of vehicles $Z$ constant. This allowed us to derive several other vehicle groupings, including $x=60$, $y=5$; $x=36$, $y=9$; $x=30$, $y=11$; $x=24$, $y=14$; $x=20$, $y=17$, among others. We have only listed the above seven grouping schemes, which are indicative of the changing trend. Similarly, we set $y=10$, $x=60$, $Z=660$; $y=12$, $x=84$, $Z=1092$, and are able to derive seven different vehicle groups. Our findings revealed that the communication complexity of B2UHChain can be reduced by up to 98.36% when compared to the single-layer blockchain, as illustrated in Fig. 7 (b). The horizontal axis of the figure represents 7 different vehicle grouping schemes, of which the fourth scheme is B2UHChain. This outcome provides evidence that the derivation of equations (10) and (11) is accurate.

*C. IoV Scalability*

In this part, we investigate the scalability of the IoV using a hierarchical blockchain approach, specifically when both the B2UHChain and single-layer blockchain have the same communication complexity in PBFT consensus. We maintain the same communication complexity as that of the previous single-layer blockchain after layering. Consequently, it was possible to increase the total number of vehicles in the system to the power of $w$, namely

$$Z^2 = 1.89(Z^w)^{\frac{4}{3}} \qquad (22)$$

Thus, we can find

$$w = 1.5 - \frac{0.48}{\lg Z} \qquad (23)$$

When the number of vehicles $Z$ is large enough, we have

$$w \approx 1.5 \qquad (24)$$

Therefore, the scalability of the B2UHChain is 1.5 power of the single-layer blockchain.

Further, similar to the calculation method of $I$, to verify consensus security after IoV expansion, we define the consensus success rate change between the double layer after capacity expansion and that before as $I'$, and the consensus success rate change between the double layer after capacity expansion and the original single layer as $I''$.

To illustrate, let us consider $Z=360$. From equation (23), it is evident that if the same consensus communication complexity is maintained as that of the single-layer blockchain, the number of vehicles in the B2UHChain can be increased to 4265. As a result, we may set $x'=200$, $y'=19$, and $Z'=4000$, as the total number of vehicles at this point is the closest to 4265, while still satisfying equation (11). We can also apply a similar expansion scheme to the B2UHChain when the single-layer blockchain has $Z$ values of 660 and 1092.

Based on the simulation results presented in Figs. 7 (c) and (d), we can observe that the security of the B2UHChain is enhanced after capacity expansion. Moreover, the negative impact on security caused by vehicle layering, as depicted in Fig. 6 (a), can be mitigated through vehicle capacity expansion.

*D. Fog Stability*

Fog stability is also a crucial performance metric. If the fog composition changes frequently, newly added vehicles must re-establish communication channels, which can adversely affect the efficiency of vehicle registration and authentication in the blockchain, and increase communication overhead. To assess the stability of fog over a certain period of driving, we define the updated rate $U$ of vehicles within fog as a critical parameter. A lower updated rate indicates better fog stability, while a higher updated rate implies the opposite.

$$U = \begin{cases} \dfrac{UV}{IV} = \dfrac{LV + NV}{IV}, & \text{Fog head vehicle have not changed} \\ 1, & \text{Fog head vehicle have changed} \end{cases} \qquad (25)$$

where $IV$ indicates the initial number of vehicles; $UV$ represents the number of vehicles updated; $LV$ indicates the number of vehicles left, and $NV$ represents the number of new vehicles added. And when the FV changes, we need to re-select FV with the rule of Section IV, therefore $U$ is 1 in this situation.

By comparing two schemes [42] and [43] with the random vehicle flow, it is proved that the vehicle grouping method in B2UHChain has the advantage of stability even in the case of the lowest communication complexity. In this simulation, we ser $Z=660$, $l=0.5$, and $m=0.5$.

First, we compare B2UHChain with the ID-MAP method [42] to find a reasonable weight allocation scheme for parameters $a$, $b$, and $c$. We establish the following allocation schemes: Scheme1: $a=0.5$, $b=0.3$, $c=0.2$; Scheme2: $a=0.5$, $b=0.2$, $c=0.3$; Scheme3: $a=0.3$, $b=0.5$, $c=0.2$; Scheme4: $a=0.3$, $b=0.2$, $c=0.5$; Scheme5: $a=0.4$, $b=0.3$, $c=0.3$; Scheme6: $a=0.3$, $b=0.4$, $c=0.3$; Scheme7: $a=0.3$, $b=0.3$, $c=0.4$. Next, we set the driving of vehicles in SUMO to be random, and determined the average updated rate of the 60 fogs to obtain the simulation results shown in Fig.8 (a).

Fig. 8 (a) indicates that Scheme 1-4 outperforms ID-MAP

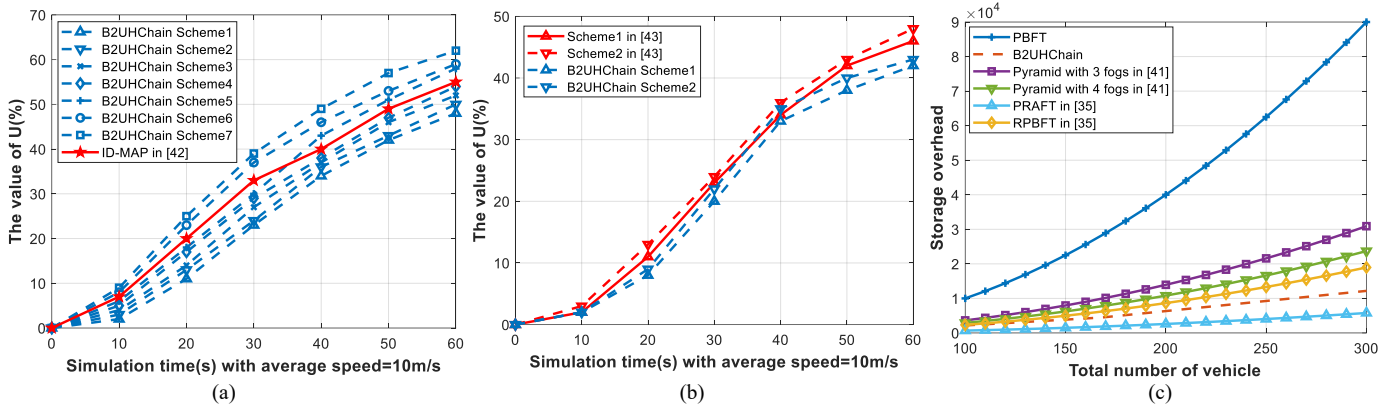
Fig. 8. (a) The fog updated rate of B2UHChain and [42]. (b) The fog updated rate of B2UHChain and [43]. (c) Storage overhead.

in terms of stability as they exhibit a lower update rate. Among them, Scheme 1 and 2 with the best stability are further selected and compared with the method in [43] using the same weight allocation scheme. The simulation results are presented in Fig. 8 (b). It is evident that the updated rate of vehicles within the fog in our scheme is slightly better than that of [43], whatever the weight allocation. This outcome suggests that fog stability is advantageous when the communication complexity of vehicle grouping is at its lowest.

The underlying principle is straightforward. In ESIA, each vehicle prioritizes the number of vehicles within the fog while calculating (13) to determine the most suitable fog. This means that in the scheme proposed in [43], if the value of $\beta_i$ in (13) changes, the vehicle may replace the fog, thereby destabilizing the composition of the fog. However, in our scheme, the number of vehicles within each fog is also taken into consideration, providing an additional constraint. Even if the value of $\beta_i$ in (13) changes, the vehicle will not replace the fog due to the limitation of the number of vehicles within each fog.

### E. Storage Overhead

Storing information in a distributed manner within a blockchain necessitates that each vehicle in the IoV stores the information of all vehicles from the same blockchain. Therefore, dividing the IoV into different fogs can also reduce the storage overhead. The storage overhead simulation for B2UHChain is similar to that of [35].

We conducted a simulation to compare the storage overhead of B2UHChain with PBFT, Pyramid in [41], PRAFT, andRPBFT in [35] simultaneously, as shown in Fig. 8 (c). The results indicate that B2UHChain has a significant advantage in storage overhead, which is substantially better than single-layer PBFT. Additionally, B2UHChain consumes less storage overhead than the two cases in Pyramid. Moreover, B2UHChain outperforms RPBFT and is only weaker than PRAFT in [35].

### F. Computation and Communication Overhead

In this part, we compare the computation and communication overhead of the ESIA scheme with that of another two IoV-oriented authentication schemes based on blockchain, which are P4C [38], and MDPA [44].

TABLE II
COMPARATIVE ANALYSIS OF COMPUTATION OVERHEAD

| Scheme | Signature (Times) | Verification (Times) | Hash operation (Times) | Encryption/ Decryption (Times) |
|---|---|---|---|---|
| P4C [38] | 4 | 2 | 4 | 4 |
| MDPA [44] | 1 | 1 | 3 | - |
| ESIA | 1 | 2 | 2 | - |

TABLE III
COMPARATIVE ANALYSIS OF COMMUNICATION OVERHEAD

| Scheme | Registration (Bytes) | Authentication (Bytes) | Total (Bytes) |
|---|---|---|---|
| P4C [38] | 144 | 660 | 804 |
| MDPA [44] | 144 | more than 440 | more than 584 |
| ESIA | 164 | 320 | 484 |

For the **computation overhead**, it includes the signature, verification, hash operation, and encryption/decryption. We have calculated the computation overhead required by these three schemes after two vehicles mutual authentication, and the comparison results are shown in Table II. Our findings show that ESIA outperforms P4C in terms of the number of signatures and encryption/decryption, while it has an advantage over MDPA in the number of hash operations. We attribute the poor performance of P4C to its focus on authentication between the vehicle and the base station. As a result, when establishing a secure channel between two vehicles, both vehicles must complete authentication with the base station.

For the **communication overhead**, we divide it into registration and authentication phases to analyze. Inspired by [44], we set the size of the blind factor, and the output of ECC are 32 bytes and 64 bytes, respectively. Moreover, the general hash output is 32 bytes, and the other data size is 4 bytes, such as the timestamp. For the registration phase of ESIA, it needs to transmit the *ReqReg*(*OID*, *FogID*, *RSUID*, *IDCard$_{FogID}$*, *Time*) message, thus, the length equals (32+32+32 +64+4)=164 bytes. Furthermore, the authentication phase of ESIA needs to transmit the *ReqCon*(*A1$_{OID}$*, *A1$_{FogID}$*, *A1$_{RSUID}$*, *A2$_{OID}$*, *IDCard$_{A1}$*), and two *IDCards* to verify, thus, the length of this phase equals (32+32+32+32+64+64+64)=320 bytes. Similarly, we also compare ESIA with P4C and MDPA, as shown in Table III. Specifically, our results show that ESIA transmits the least number of bytes, and has the lowest communication overhead.

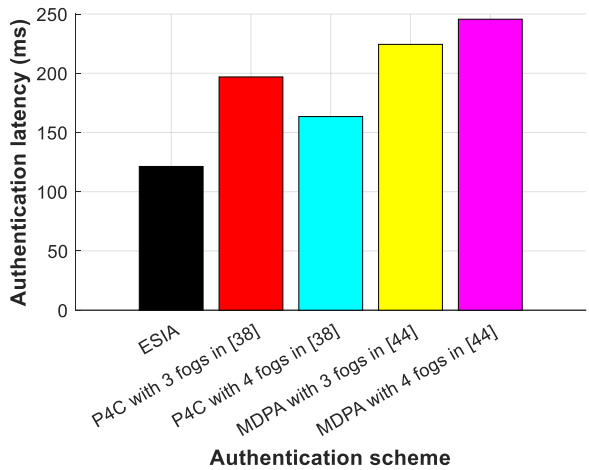

Fig. 9. Authentication latency.

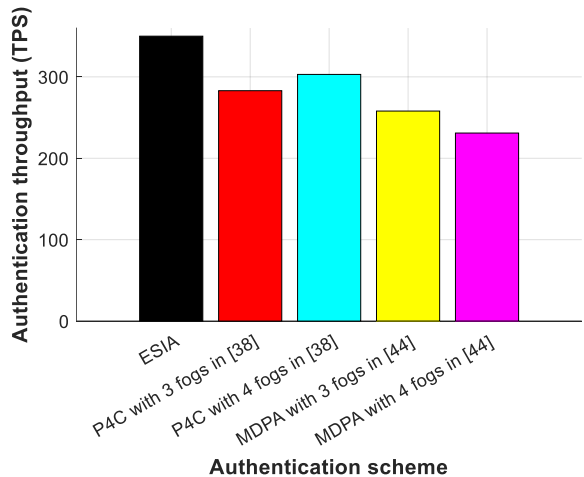

Fig. 10. Authentication throughput.

### G. Latency and Throughput

In this part, we evaluate the latency and throughput of our ESIA under the BU2HChain structure, and compare that with the above P4C [38] and MDPA [44], when two vehicles have done the mutual authentication. Our experimental setup involved using Python 3.7 on a PC equipped with an Intel I7-1260P processor, which has a clock frequency of 2.1 GHz and 16 GB RAM. And we set the number of vehicles is 100.

Fig. 9 presents the authentication latency comparison between ESIA, P4C, and MDPA. To evaluate the performance of P4C and MDPA schemes under different fog numbers, we simulate their performance with 3 and 4 fogs, respectively. Our results indicate that ESIA outperforms both P4C and MDPA in terms of authentication latency, with a latency of only 121 ms. Regarding P4C, we observe that increasing the number of fogs from 3 to 4 can reduce the authentication latency due to the DPBFT consensus's lower communication complexity under a higher number of fogs. However, for MDPA, increasing the number of fogs led to an increase in communication rounds during the consensus, resulting in higher authentication latency.

Fig. 10 illustrates the authentication throughput comparison between ESIA, P4C, and MDPA. Similar to the previous part, we simulate the performance of P4C and MDPA with 3 and 4 fogs. The throughput in this simulation is the maximum measured at different authentication request rates. Our results reveal that ESIA has superior authentication throughput, achieving up to 350 transactions per second (TPS). However, we note that the throughput performance of P4C and MDPA exhibits a different behavior than that of the latency under varying fog numbers, but the principle is consistent with the latency part and is not further discussed here.

### VIII. CONCLUSION

In this paper, we have introduced vehicular fog computing to design a hierarchical blockchain called B2UHChain for IoV, with the objective of minimizing the communication complexity of consensus and ensuring the stability of vehicular fog. Based on this structure, we have designed an efficient and stable identity authentication scheme called ESIA.

Theoretical analysis and simulation results demonstrate that the B2UHChain offers higher scalability compared to single-layer blockchains, and can increase the capacity of IoV by a power of 1.5. Additionally, B2UHChain outperforms other hierarchical blockchains in terms of communication complexity, consensus security, stability, and storage overhead. Furthermore, our proposed authentication scheme, ESIA, offers advantages over other authentication schemes in terms of communication and computing overhead, latency, and throughput. In the future, we plan to explore the application of B2UHChain in other network scenarios.